# Investigating the Temperature Effects on Resistive Random Access Memory (RRAM) Devices

T. D. Dongale [1, †], K. V. Khot [2], S. V. Mohite [3], S. S. Khandagale [d], S. S. Shinde [3], A. V. Moholkar [3], K. Y. Rajpure [3], P. N. Bhosale [2], P. S. Patil [3], P. K. Gaikwad [4], R. K. Kamat [4]

[1] Computational Electronics and Nanoscience Research Laboratory,
School of Nanoscience and Biotechnology, Shivaji University Kolhapur- 416004
[2] Department of Chemistry, Shivaji University, Kolhapur 416004, India
[3] Department of Physics, Shivaji University, Kolhapur 416004, India
[4] Embedded System and VLSI Research Laboratory, Department of Electronics,
Shivaji University, Kolhapur, 416004, India

**Abstract**

In this paper, we report the effect of filament radius and filament resistivity on the saturated temperature of ZnO, $TiO_2$, $WO_3$ and $HfO_2$ Resistive Random Access Memory (RRAM) devices. We resort to the thermal reaction model of RRAM for the present analysis. The results substantiate decrease in saturated temperature with increase in the radius and resistivity of filament for the investigated RRAM devices. Moreover, a sudden change in the saturated temperature at a lower value of filament radius and resistivity is observed as against the steady change at the medium and higher value of the filament radius and resistivity. Results confirm the dependence of saturated temperature on the filament size and resistivity in RRAM.

† **Corresponding Author:** T. D. Dongale
E-mail: tdd.snst@unishivaji.ac.in

# 1. Introduction

The Resistive Random Access Memory (RRAM) is considered to be a strong candidate which is poised to substitute the conventional flash memories [1]. The RRAM have several advantages over its other counterparts, for instance, high density of data storage, long data retention, low operating voltage, high endurance, fast switching speed and compatibility with conventional CMOS process and so on [1-2]. Generally, physical and chemical mechanism of RRAM is classified in terms of Valency Change Mechanism (VCM) [3-5], Electro-Chemical Metallization (ECM) [6], Thermo-Chemical Mechanism (TCM) [7], Phase Change Mechanism (PCM) [8], and Electrostatic/Electronic Mechanism (EEM) [9] etc. In above all mechanism the resistance of the RRAM switches between two resistance states viz. low resistance state (LRS) and high resistance state (HRS) [10-11]. The popular version of RRAM is known as a memristor which is modeled around the valency change mechanism [12-15]. Our group has worked to a greater depth in the area of design, development, modeling, realization and applications of memristors [3-5, 10-19].

Temperature plays an important role in resistive switching and many studies have been devoted to finding out the underlying physical mechanism and its effect on memory performance. Recently Shang et al reported the heterostructure based thermally stable transparent RRAM. The reported device exhibits forming-free bipolar resistive switching behavior at room temperature with good memory performance [20]. Wang et al reported the thermoelectric Seebeck effect in oxide-based resistive switching memory. They have investigated the intrinsic electronic transport mechanism by measuring thermoelectric Seebeck effects [21]. Yi et al reported the effect of annealing temperature on graphene oxide-based RRAM. Their results are evident that the lower annealing temperature improves the memory performance of graphene oxide-based RRAM [22]. Fang et al reported low-temperature switching characteristics and conduction mechanism of $HfO_x$ based RRAM. The results suggested that at a lower temperature the switching voltage increases [23]. Tsuruoka et al reported the effect of temperature on Cu–$Ta_2O_5$-based atomic switch and investigated the switching mechanism of the developed device. The results revel that SET and RESET voltage decreases as the temperature increases [24]. Sato et al reported the thermal reaction model of the metal oxide-based RRAM. This model calculates the temperature of the conductive filament and its corresponding effect on RRAM performance [25].

In light of the literature review and as an extension of our ongoing work [3-5, 10-19] the present manuscript, reports our investigations on the effect of temperature on ZnO, TiO$_2$, WO$_3$ and HfO$_2$ based RRAM devices using thermal reaction model reported in the ref. [25]. The rest of the paper is as follows, after a brief introduction in the first section, the second section deals with the overview of thermal reaction model of RRAM and other computational details. The third section deals with the effect of filament radius and resistivity on the saturated temperature of ZnO, TiO$_2$, WO$_3$ and HfO$_2$ based RRAM devices. At the end conclusion is portrayed.

## 2. Thermal Reaction Model of RRAM

The temperature of the conductive filament plays an important role in the resistive switching. It is observed that the high current density present in the conductive filament and the aftermaths of temperature effects are unavoidable [26]. Hence, the analysis of temperature effect is of foremost important. Fig. 1 represents the cross-sectional view of the thermal reaction model [25].

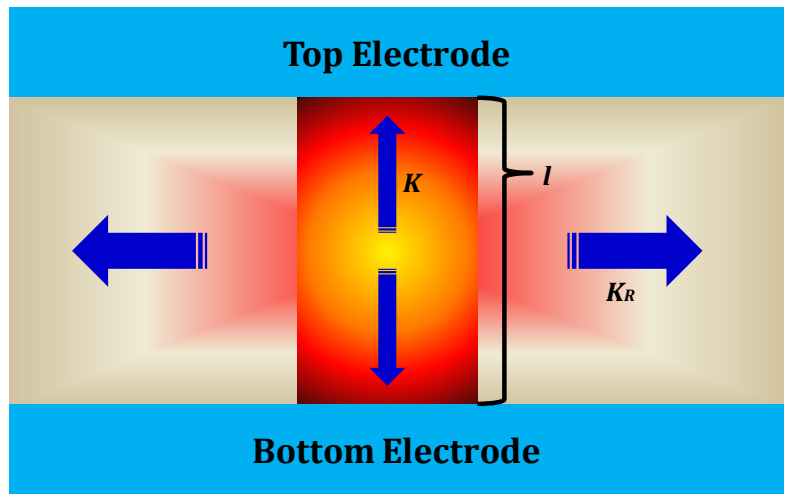

**Fig. 1:** Cross sectional view of the model under analysis. This model is popularly known as thermal reaction model of RRAM Device [25].

The model assumes the cylindrical conducting path present in the active layer (oxidation membrane) of the RRAM. The heating temperature $\Delta T$ can be represented as [25],

$$\Delta T = \frac{1}{(K_R + K)} \frac{V_{reset}^2}{R} \tag{1}$$

where, $V_{reset}$ is the RESET voltage of RRAM, $R$ is a resistivity of the filament, $K_R$ is known as the radial thermal conductance of oxidation membrane and $K$ is the filament thermal conductance and they are given as [25],

$$K_R = \frac{2\pi k' l/4}{\log_e (r_2/r_1)} \quad (2)$$

$$K = \frac{k\pi r_1^2}{l/2} \quad (3)$$

where $k'$ and $k$ are the thermal conductivities of the materials, $l$ is a thickness of oxidation membrane, and $r$ is the radius of the conductive filament. Table 1 represents the various physical parameters for simulation of different RRAM structures.

**Table 1:** Simulation parameters for various RRAM devices

| TMO materials | Thermal conductivity-k' (W/cm/ °C) | Thermal conductivity-k (W/cm/ °C) | Specific heat-c' (J/g/ °C) | Density of material-γ' (g/cm³) |
|---|---|---|---|---|
| ZnO | 1.2 | 116 | 0.5 | 5.61 |
| TiO$_2$ | 6.69 | 21.9 | 0.6894 | 4.13 |
| WO$_3$ | 1.63 | 173 | 0.0780 | 7.16 |
| HfO$_2$ | 1.1 | 23 | 0.144 | 13.31 |

In the present analysis thickness of the oxidation membrane is kept constant at 200 nm, and reset voltage at 0.5 V. To analyze the saturated temperature behavior of different RRAM devices, we have varied the filament radius in the range of 10 nm to 100 nm with 10 nm as a step size and filament resistivity in the range of 10 μΩ cm to 100 μΩ cm with 10 μΩ cm as a step size. To analyze the effect of filament radius on different RRAM devices, we have taken three observation of filament resistivity in the order of 10 μΩ cm (low resistivity), 50 μΩ cm (medium resistivity), and 100 μΩ cm (high resistivity). Similarly, analyzing the effect of filament resistivity on different RRAM devices, we have chosen three observations for filament radius size in the order of 10 nm (low size), 50 nm (medium size), and 100 nm (high size).

## 3. Result and Discussion

In the present investigation, different RRAM materials are analyzed such as ZnO, TiO$_2$, WO$_3$ and HfO$_2$ owing to their fine memory performance [3-5]. The various physical parameters of above RRAM materials are listed in table 1. For the first case, the radius of the filament is varied from 10 nm to 100 nm and its effect on the saturated temperature of different RRAM material is analyzed. For each case, the resistivity of the filament is changed as 10 μΩ cm, 50 μΩ cm, and 100 μΩ cm. Fig 2 (a) represent the effect of filament radius on the saturated temperature of ZnO based RRAM device. The results suggest that as the radius of the filament increases the corresponding saturated temperature decreases. It is also seen that saturated temperature decreases as the resistivity of the filament increases.

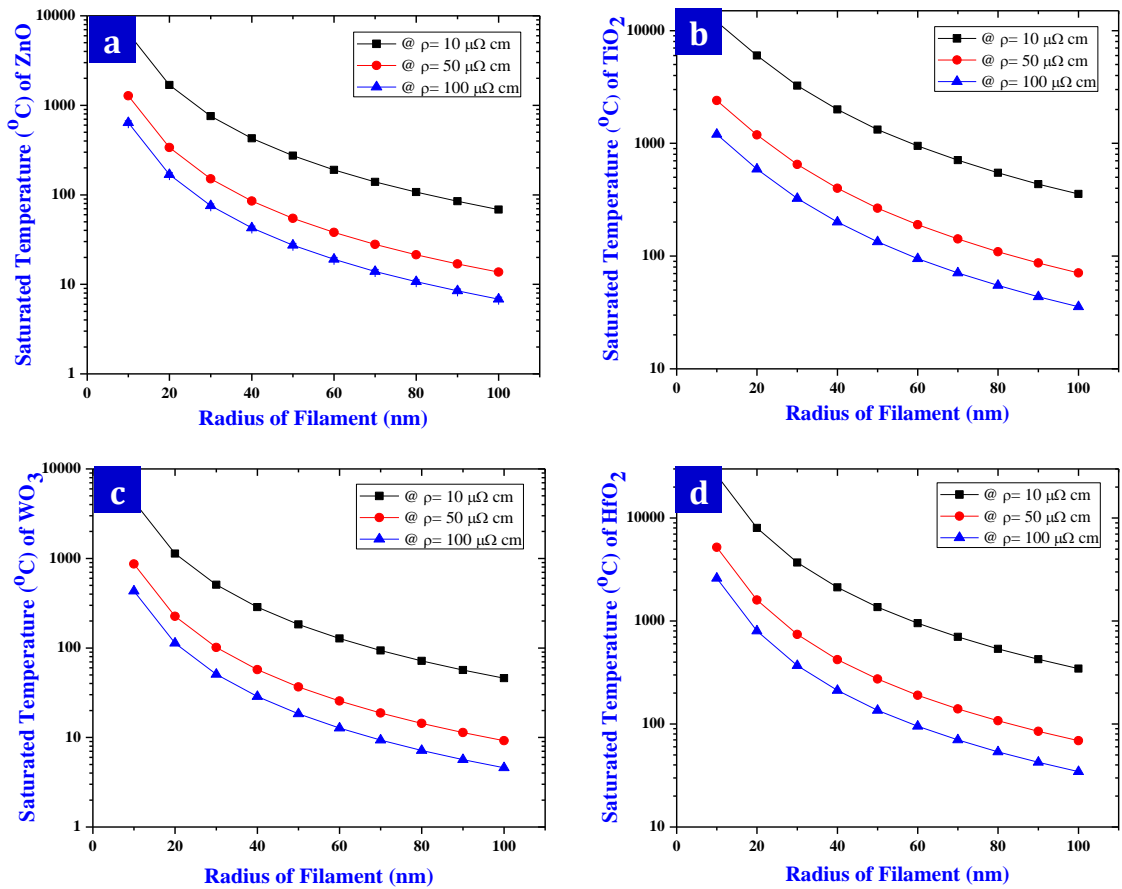

**Fig. 2:** Effect of the radius of the filament on the saturated temperature of (a) ZnO; (b) TiO$_2$; (c) WO$_3$; and (d) HfO$_2$ based RRAM.

Fig 2 (b to d) present the effect of filament radius on the saturated temperature of TiO$_2$, WO$_3$, and HfO$_2$ based RRAM devices. The results clearly show the inverse relationship between the radius of the filament and saturated temperature for each RRAM devices. The results also prompt that WO$_3$ based RRAM device has a lower value of saturated temperature than its other counterparts for the same value of radius and resistivity of the conducting filament. The HfO$_2$ based RRAM device shows the higher value of saturated temperature than other RRAM materials for the same value of radius and resistivity of the conducting filament. The results also indicate that there is a steady change in the saturated temperature value at medium (50 µΩ cm) and higher (100 µΩ cm) value of the filament resistivity, but sudden change is observed at the lower (10 µΩ cm) value of the filament resistivity. The same may be attributed to the lower value of filament resistivity invoking a high current density conductive filament that gives rise to higher heat dissipation.

For the second case, the resistivity of the filament is varied in the range of 10 µΩ cm to 100 µΩ cm with 10 µΩ cm as a step size and its effect on the saturated temperature of different RRAM material is analyzed. For each case, the filament radius is changed as 10 nm, 50 nm, and 100 nm. Fig 3 (a) represent the effect of filament resistivity on the saturated temperature of ZnO based RRAM. The results suggest that saturated temperature decrease with increase in filament resistivity. It is also seen that saturated temperature decreases as the radius of the filament increases from 10 nm to 100 nm.

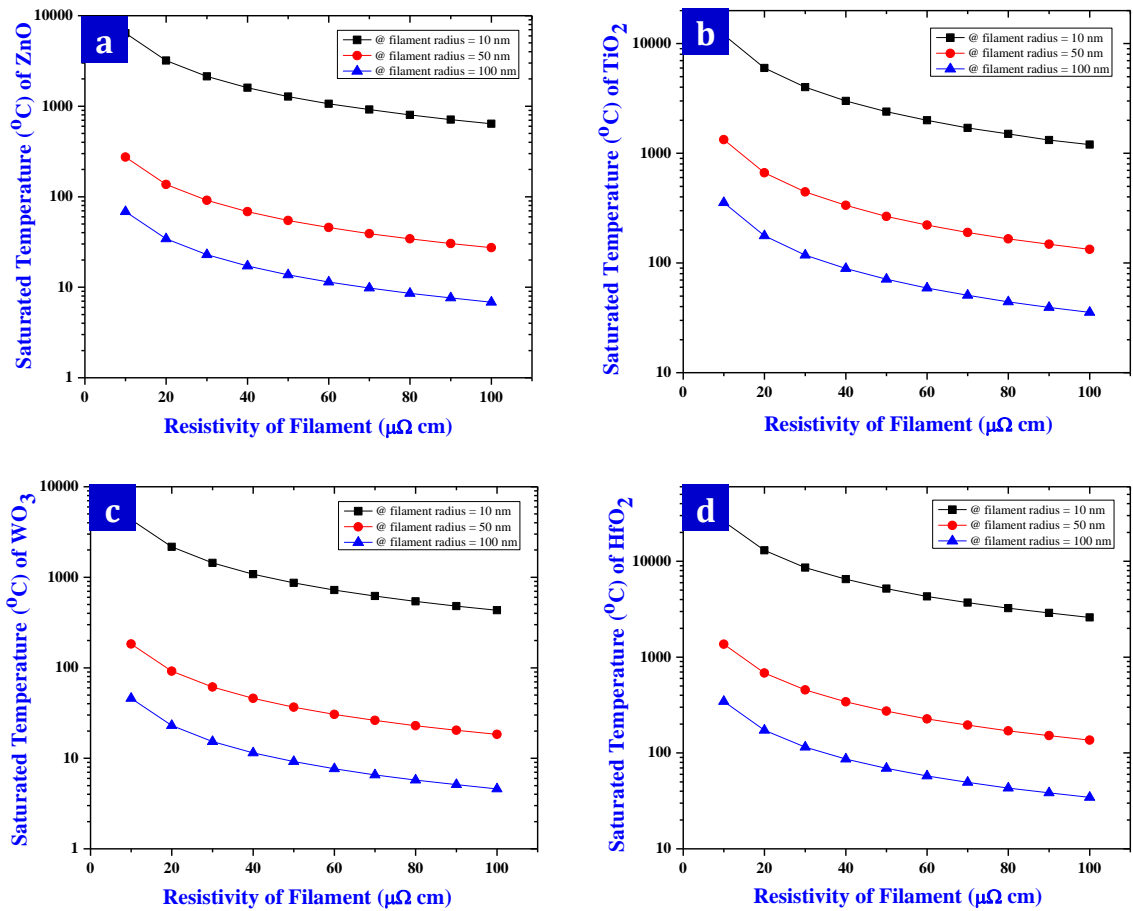

**Fig. 3:** Effect of the resistivity of the filament on the saturated temperature of (a) ZnO; (b) TiO$_2$; (c) WO$_3$; and (d) HfO$_2$ based RRAM.

Fig 3 (b to d) represent the effect of filament resistivity on saturated temperature for TiO$_2$, WO$_3$, and HfO$_2$ based RRAM devices. Here too an inverse relationship is observed between the resistivity of the filament and saturated temperature. The results indicate that WO$_3$ based RRAM device has a lower value of saturated temperate than its other counterparts for the same value of filament resistivity and radius. The HfO$_2$ based RRAM device shows the higher value of saturated temperature than other RRAM materials for the same value of filament resistivity and radius.

Table 2 and 3 represents the saturated temperature change factor for the change in the filament resistivity and radius of different materials. The results prompt that, the rate of change in the saturated temperature is higher for the lower resistivity whereas it becomes small for higher filament resistivity. The variation in the filament radius shows same behavior for ZnO, WO$_3$ and HfO$_2$ based RRAM devices however TiO$_2$ material shows a small variation in the change factor. This is attributed to the higher thermal conductivity and specific heat capacity of TiO$_2$ as compared to other ones taken in this investigation. The results showcase that the saturated temperature is the filament size and resistivity dependent property.

**Table 2:** Saturated temperature change factor for the change in the filament resistivity value.

| RRAM Material | Saturated Temperature Change Factor for 10 µΩ cm to 50 µΩ cm. | Saturated Temperature Change Factor for 50 µΩ cm to 100 µΩ cm. |
|---|---|---|
| ZnO | 5.0022 | 1.9997 |
| TiO$_2$ | 5.0050 | 1.9983 |
| WO$_3$ | 5.0006 | 1.9998 |
| HfO$_2$ | 5.0031 | 1.9994 |

**Table 3:** Saturated temperature change factor for the change in the filament radius value.

| RRAM Material | Saturated Temperature Change Factor for 10 nm to 50 nm. | Saturated Temperature Change Factor for 50 nm to 100 nm. |
|---|---|---|
| ZnO | 23.3900 | 3.9908 |
| TiO$_2$ | 8.9945 | 3.7538 |
| WO$_3$ | 23.5772 | 3.9927 |
| HfO$_2$ | 19.0357 | 3.9540 |

## 4. Conclusion

In the present investigation, we have thoroughly investigated the thermal reaction model of RRAM device for different material and consequently correlated the relationship between saturated temperature, filament radius, and resistivity. The saturated temperature is one of the important property of high-performance RRAM and it is associated with the device reliability. The higher value of the saturated temperature degrades the device reliability [26]. At the higher temperature more sensitive detection circuitry is required to read and write operations of RRAM [5, 25-26]. The results suggest that the saturated temperature decreases as the radius and resistivity of the conductive filament increases. The results also indicate that there is a sudden change in the saturated temperature for the lower value of filament radius and resistivity whereas steady change is observed for the medium and higher value of the filament radius and resistivity. Results confirm the saturated temperature is filament size and resistivity dependent property. All these investigations and results thereof are definitely significant in the application scenario of RRAMs.